\documentclass[manuscript]{aastex}
\usepackage{epstopdf,natbib}

\shorttitle{The Yarkovsky Drift's Influence on NEAs}
\shortauthors{Nugent et al.}
\begin{document}


\title{The Yarkovsky Drift's Influence on NEAs: Trends and Predictions with NEOWISE Measurements}


\author{C. R. Nugent\altaffilmark{1}, A. Mainzer\altaffilmark{2},  J. Masiero\altaffilmark{2},
T. Grav\altaffilmark{3}, and J. Bauer\altaffilmark{2}}


\altaffiltext{1}{Department of Earth and Space Sciences, University of California, Los Angeles, CA 90095, USA}
\altaffiltext{2}{Jet Propulsion Laboratory, California Institute of Technology, Pasadena, CA 91109 USA}
\altaffiltext{3}{Planetary Science Institute, Tucson, AZ}


\begin{abstract}
We used WISE-derived geometric albedos ($p_V$) and diameters, as well as geometric albedos and 
diameters from the literature, to produce more accurate diurnal Yarkovsky drift 
predictions for 540 near-Earth asteroids (NEAs) out of the current sample of $\sim8800$ known objects.  As 
ten of the twelve objects with the fastest predicted rates have observed arcs of 
less than a decade, we list upcoming apparitions of these NEAs to facilitate observations. 
\end{abstract}






\section{Introduction}


The Yarkovsky effect is a non-gravitational force that perturbs the orbits of small 
bodies, including near-Earth asteroids (NEAs). Despite its small magnitude, it must 
be included in the calculation of precise asteroid trajectory predictions \citep
{Giorgini1950DA,MilaniImpact1999RQ36}, and it is believed to be a key mechanism in 
the process that delivers asteroids from the main belt to near-Earth space \citep
{2006AREPS}.

The diurnal Yarkovsky effect (or drift) is caused by anisotropic re-radiation of absorbed 
sunlight. It is driven by the thermal properties of an asteroid as well as the 
amount of absorbed incident radiation. A given surface point on an asteroid 
observes maximum incident radiation at local noon, but thermal inertia causes the 
time of maximum emitted radiation (usually at infrared wavelengths) to occur later. 
Each arriving and departing photon has an associated momentum $p = E/c$, where $E$ is 
the photon's energy and $c$ is the speed of light. Since the body is rotating, the 
incident radiation is in a different direction than the later emitted radiation, 
and the body experiences a very small net acceleration. If the body has a prograde 
spin, the net acceleration has a component aligned with the motion of the body's 
orbit, nudging the body away from the sun. Similarly, a body with a retrograde spin 
will feel an acceleration with a component anti-aligned with its velocity, shifting 
it towards the sun \citep{2006AREPS}.

There is also a seasonal component to the Yarkovsky effect. The seasonal Yarkovsky effect is largest when an asteroid's obliquity is $90^\circ$, and goes to zero as obliquity approaches $0^\circ$ or $180^\circ$ \citep{2006AREPS}. \citet{VokSmallNEAS} calculated the diurnal and seasonal components of the Yarkovsky effect for several objects, and in all cases the seasonal component was significantly smaller. Even in the case of (1566) Icarus, which has an obliquity equal to $103^\circ$, the diurnal component for this object was more than twice the magnitude of the seasonal component over a range of likely thermal conductivities.

There have been few direct measurements of the Yarkovsky drift. \citet
{ChesleyGolevka} used radar ranging to make the first direct detection of the 
Yarkovsky drift. They measured the rate of change of (6489) Golevka's semi-major 
axis ($da/dt$) to be of order $10^{-4}$ AU/Myr. A magnitude $da/dt$ of $10^{-3}$ 
AU/Myr Yarkovsky drift was associated with asteroid 1992 BF by linking modern 
astrometry with observations from 1952 \citep{Vok1992BF}. \citet{Nugent12} used an 
 orbit-fitting method to measure Yarkovsky drifts for 54 NEAs, and found an average rate magnitude of $(10.4 \pm 11.4) \times 10^{-4} $ AU/Myr. 

The Yarkovsky effect has been modeled by several researchers (\citet
{VokSmallNEAS, SpitaleGreenberg01}, for example). Mathematical formulations, such 
as those by \citet{VokSmallNEAS}, indicate that Yarkovsky drift 
 is inversely proportional to diameter. 
Although the amount of absorbed radiation increases with the square of the 
diameter, mass increases with the cube of the diameter ($D$), so drift rate is expected 
to show a $1/D$ dependence. 

However, because thermal inertia could also depend on size, the size-dependence of 
Yarkovsky drift could be more complicated than presently assumed. Theory predicts 
that the more massive a body is, the more regolith it should retain \citep
{ScheeresFateEjecta}, and regolith may act as an insulating blanket  
 (though for bodies smaller than 10 km in diameter, spin state may be more indicative 
 of regolith presence). Low 
porosity and high thermal inertia should create a longer time lag between absorbed 
radiation and thermal re-radiation, perhaps resulting in a stronger Yarkovsky 
effect (depending on the rotation state).

Additionally, these models incorporate physical properties of asteroids that are 
often poorly measured. Although obliquity, heat capacity, thermal conductivity, and 
bulk density are generally difficult to quantify, more basic properties such as 
geometric albedo ($p_V$) and diameter can be ill-constrained. This dearth of information has hindered 
the accuracy of Yarkovsky predictions.

The Wide-field Infrared Survey Explorer (WISE) \citep{Wright2010} has observed over 150,000 minor planets (including $\sim600$ NEAS) at infrared 
wavelengths \citep{Mainzer2011a}. It is these infrared measurements, combined with 
optical observations, that can separate the contributions of size and $p_V$ to the 
observed flux. The dependence of flux on $p_V$ is weaker in the thermal 
wavelengths, since the majority of the light emitted from the asteroid is from 
thermal emission, not reflected infrared sunlight. With a thermal model that 
incorporates both infrared and optical observations, size and $p_V$ can be 
determined. With thermally-dominated WISE wavelengths, it has been shown that for asteroids observed with good 
signal-to-noise ratios and relatively low amplitude lightcurve variations,  diameter 
can be determined to within $\pm 10\%$, and $p_V$ can be determined to within $\pm 
25\%$ of the amount of the albedo \citep{Mainzer2011b,Mainzer2011c}. Combining WISE measurements with published reliable 
diameters determined from in situ spacecraft visits, 
stellar occultations, and radar produces a list of NEOs with well-determined 
diameters and geometric albedos.

\section{Methods}
\label{sec:methods}

We employed the mathematical formulation of the diurnal Yarkovsky effect developed 
by \citet{VokSmallNEAS} to numerically estimate Yarkovsky drifts. Although our
methods are not identical, this work follows that of \citet{VokDetect}, who
predicted drifts for 28 NEAs. We expand from that foundation, incorporating newly
available physical properties.

For a time step along an NEA's orbit, the Yarkovsky acceleration was computed 
following equation (1) of \citet{VokSmallNEAS}. This equation assumes a spherical 
body and that temperatures throughout the body do not greatly deviate from an 
average temperature. Obliquity was assumed to be $0^\circ$ to produce maximum 
drift. Therefore, the reported drifts in this paper are upper limits. 
Additionally, a $0^\circ$ obliquity assumes that all drift is due to the diurnal
Yarkovsky effect, as the seasonal Yarkovsky effect has zero magnitude for this case
 \citep{2006AREPS}.
This acceleration 
was resolved along orthogonal directions, and Gauss' form of Lagrange's planetary 
equations \citep{Danby} was employed to evaluate an orbit-averaged $da/dt$.

The magnitude of the diurnal Yarkovsky drift depends on physical parameters which 
can be ill-defined. 
The drift magnitude is not linearly related to these unknown parameters,
 and so the resultant drift magnitude was statistically modeled to more
 accurately determine the effect of these uncertainties.
 For each NEA, we used $p_V$ and diameter measurements derived 
by WISE \citep{Mainzer2011d} or other sources of reliable diameter and $p_V$ 
measurements in the literature, primarily radar detections and stellar 
occultations. We employed a Monte Carlo method to explore how variations in 
physical parameters contribute to errors in the prediction of $da/dt$. For $1000$ 
realizations per NEA, we added Gaussian-distributed noise to the diameter and 
$p_V$ measurements, so that standard distribution of the noise corresponded to the 
$1\sigma$ error bars on those measurements. 

As the formulation of \citet{VokSmallNEAS} relies on Bond albedo $A$, we approximated A using $A\approx(0.290 + 0.684 G)p_V$, where $G$ is the slope parameter \citep{bowe89}. In seven cases, $G$ was available in the JPL Small-Bodies database \citep{SmallBodyDatabase}. In the remaining cases, $G$ was taken to equal 0.15, as this was the value used to compute physical properties of NEAs in \citet{Mainzer2011a}, based on the standard value assumed by the Minor Planet Center for computing $H$.

Additionally, we varied the thermal 
conductivity, bulk density, and density of the surface layer between the ranges 
shown in Table \ref{tab:thercon}. The physical parameters in Table \ref
{tab:thercon} were chosen to represent a range of asteroid compositions, so that 
our Yarkovsky estimates would represent reasonable estimates of the range of 
physical properties of rocky asteroids. At one end of the spectrum are physical 
parameters mimicking a low-density rubble pile, at the other, a regolith-free rock 
chunk.  

Emissivity was always assumed to be 0.9. If rotation rate was not available in the JPL 
Database \citep{SmallBodyDatabase}, the rotation rate was assumed to be 5 revolutions/day, 
based on the average spin rate values for asteroids 1 to 10 km in diameter shown in 
Figure 1 of \citet{PravecFastSlow}. Rotation rates were unavailable for $81\%$ of the NEAs.

The $da/dt$ values quoted in this paper are the mean of these 1000 realizations. 
Error bars on $da/dt$ were determined by computing the standard deviation from the 
mean.

\begin{table}[h]
\begin{center}
	\caption{Physical and thermal properties used for generating
	      predictions of $da/dt$ drifts. Thermal
          properties are based on the work of \citet{OpeilTherCon},
          who measured three meteorites at 200~K. Listed are heat
          capacity $C$, thermal conductivity $K$, bulk density of the
          surface $\rho_s$, and mean bulk density
          $\rho_b$. The surface and bulk densities are assumed to 
          have a similar range of values, however, $\rho_s$ was not 
          necessarily equal to $\rho_b$ for a given object and 
          realization.\label{tab:thercon} }
\begin{tabular}{lccc}
\tableline\tableline
 Composition  & $C$ (${\rm J\ kg}^{-1}{\rm K}^{-1}$)& $K$ (${\rm W\ m}^{-1} {\rm K}^{-1}$) & $\rho_s/\rho_b$ (${\rm kg\  m}^{-3}$ ) ) \\ 
 \tableline
Rubble Pile & 500 & 0.01 & 1000 \\ 
Rock Chunk & 500 & 0.50 & 3000 \\ 
\tableline
\end{tabular}
\end{center}
\end{table}

\section{Results}

We estimated diurnal Yarkovsky drifts for 540 NEAs with measured diameters and 
geometric albedos. The dozen objects with the highest drifts are 
listed in Table~\ref{tbl:dozen}, upcoming apparitions of those objects are in Table~\ref{tbl:obs}, and predicted drifts for all objects are in Table~\ref{tbl:all}. Tables 
\ref{tbl:dozen} and \ref{tbl:all} include an order of magnitude 
estimate of along-track displacement ($\Delta \rho$) that would result from the 
$da/dt$ drift over 10 years. For this we use the following formulation from \citet
{VokSmallNEAS},
\begin{equation}
\Delta \rho \simeq 7 \dot{a}_4 (\Delta_{10} t)^2 a_{AU}^{-3/2}
\end{equation}
where $\Delta \rho$ is in units of km, $\dot{a}_4$ is $da/dt$ in units of
$10^{-4}$ AU/Myr, $\Delta_{10} t$ is the time difference between
observations in tens of years, and $a_{AU}$ is the semimajor axis of
the object in AU.  We note that the four of the twelve objects with the largest predicted drifts were 
discovered by the NEOWISE portion of the WISE mission (2010 JG87, 2010 HX107, 2010 EX11, and 2010 GQ75).

\begin{figure}[p]
	\begin{center}
		\includegraphics[scale=1.1]{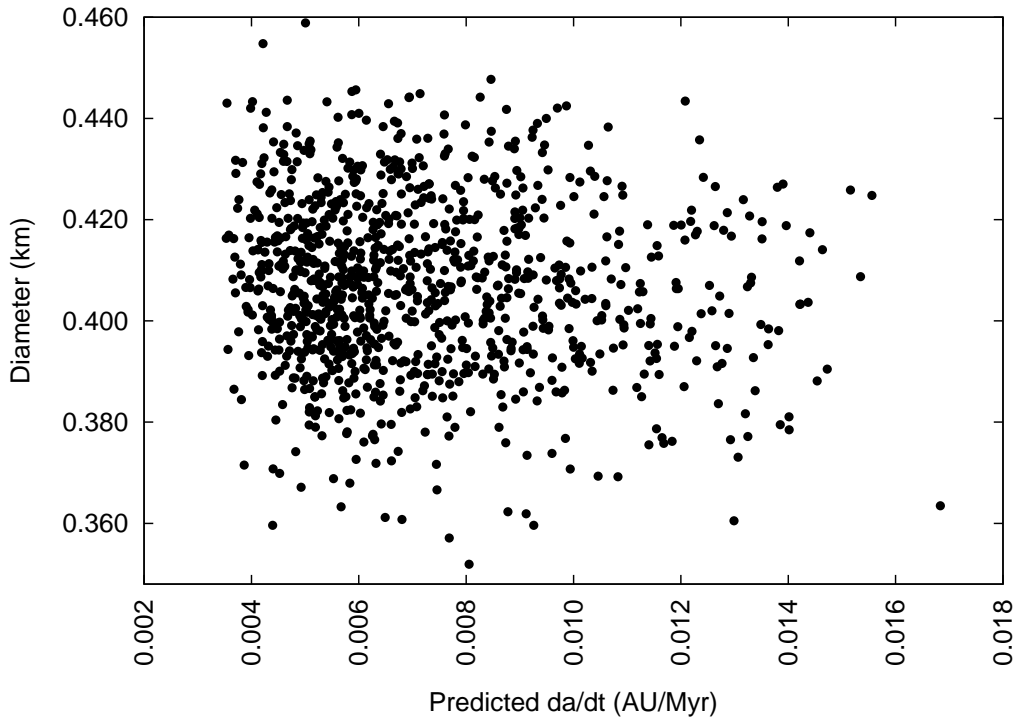}
	\end{center}
	\caption{1000 realizations of diameter vs $da/dt$ for NEA 2010 JG87. Each point 
	represents the drift produced by a different combination of physical 
	parameters. This 
	object 
	has the fastest predicted diurnal drift of all the NEAs in this paper, with 
	$da/dt = (72.11 \pm 25.12) \times 10^{-4}$ AU/Myr. Although Yarkovsky drift has a 
	$1/D$ dependence, the relatively small error bars on this object's diameter (and therefore 
	the small range of diameters shown in this plot), combined with the variations in 
	the other parameters (surface density, bulk density, thermal conductivity $K$, $p_v$, and $G$)  prevent this dependence from
	 being immediately apparent in this figure. For a clearer illustration of the 
	 relationship between $da/dt$ and diameter, see Figure \ref{fig:radsize}. For the 
	 relationship between $da/dt$ and the other physical properties that were 
	 varied during each realization, see Figure \ref{fig:EX11all}. \label
	 {fig:EX11diam}}
\end{figure}

\begin{figure}[p]
	\begin{center}
		\includegraphics[scale=1.1]{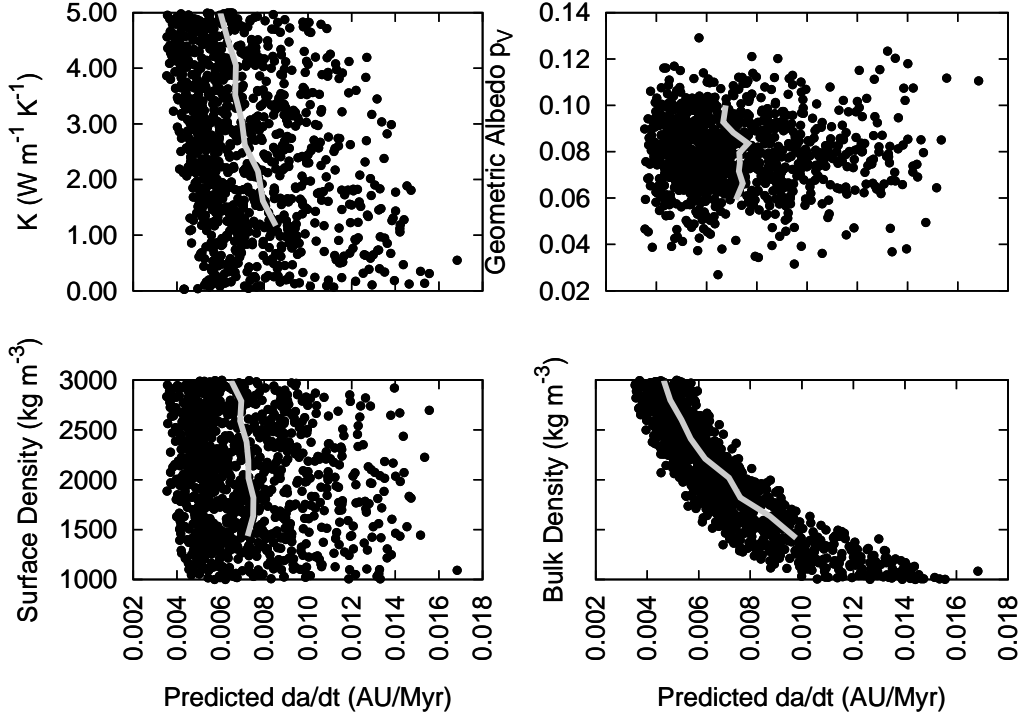}
	\end{center}
	\caption{1000 realizations of predicted diurnal $da/dt$ drift for 2010 JG87, 
	the NEA with the fastest predicted diurnal drift in this paper. For each 
	realization, diameter, thermal conductivity, geometric albedo, slope parameter $G$, density of the surface 
	layer, and bulk density were varied as described in the text. Grey lines are 
	running averages. For this object, it is the uncertainty in bulk density that 
	is mainly responsible for the span of calculated $da/dt$ drifts,
	 as the diameter of this object 
	is well-constrained (see Figure \ref{fig:EX11diam}). Also visible are the relationships between 
	thermal conductivity and surface density and drift. These two properties govern 
	the thermal lag angle.
	\label{fig:EX11all}
}
\end{figure}

Individual realizations for the NEA with the fastest predicted drift, 2010 JG87, 
are examined in Figures \ref{fig:EX11diam} and \ref{fig:EX11all}. In each of these 
figures, all 1000 realizations of physical parameter combinations are shown, so 
their individual influences are apparent for this object.

2010 JG87's diameter was determined to within $\pm10\%$ \citep{Mainzer2011d} based on the WISE 
observations (Figure \ref{fig:EX11diam}), and as diameter and bulk density are used 
to estimate mass, it is the uncertainty in bulk density that mainly determines the 
predicted drift for this object (Figure \ref{fig:EX11all}). Surface density and 
thermal conductivity both contribute to the thermal lag, and for this object, low 
values of $K$ and $\rho_s$ lead to a thermal lag that produces the strongest drifts 
(given the object's assumed rotation period of 5 revolutions/day). As geometric albedo has been determined to 
be $0.20\pm0.04$ for this object, the range of geometric albedo values explored do not 
strongly influence the resulting drift.

We now examine the values that govern Yarkovsky strength for all objects in our 
sample. As we are comparing the mean $da/dt$ values of each object, the following 
compares drifts effectively computed with the same bulk density and density of the 
surface layer. The predicted diurnal $da/dt$ has a 1/D dependence, and also 
depends on the  
amount of average incident radiation the NEA receives per orbit and $da/dt$. 

\begin{figure}[p]
	\begin{center}
		\includegraphics[scale=1.1]{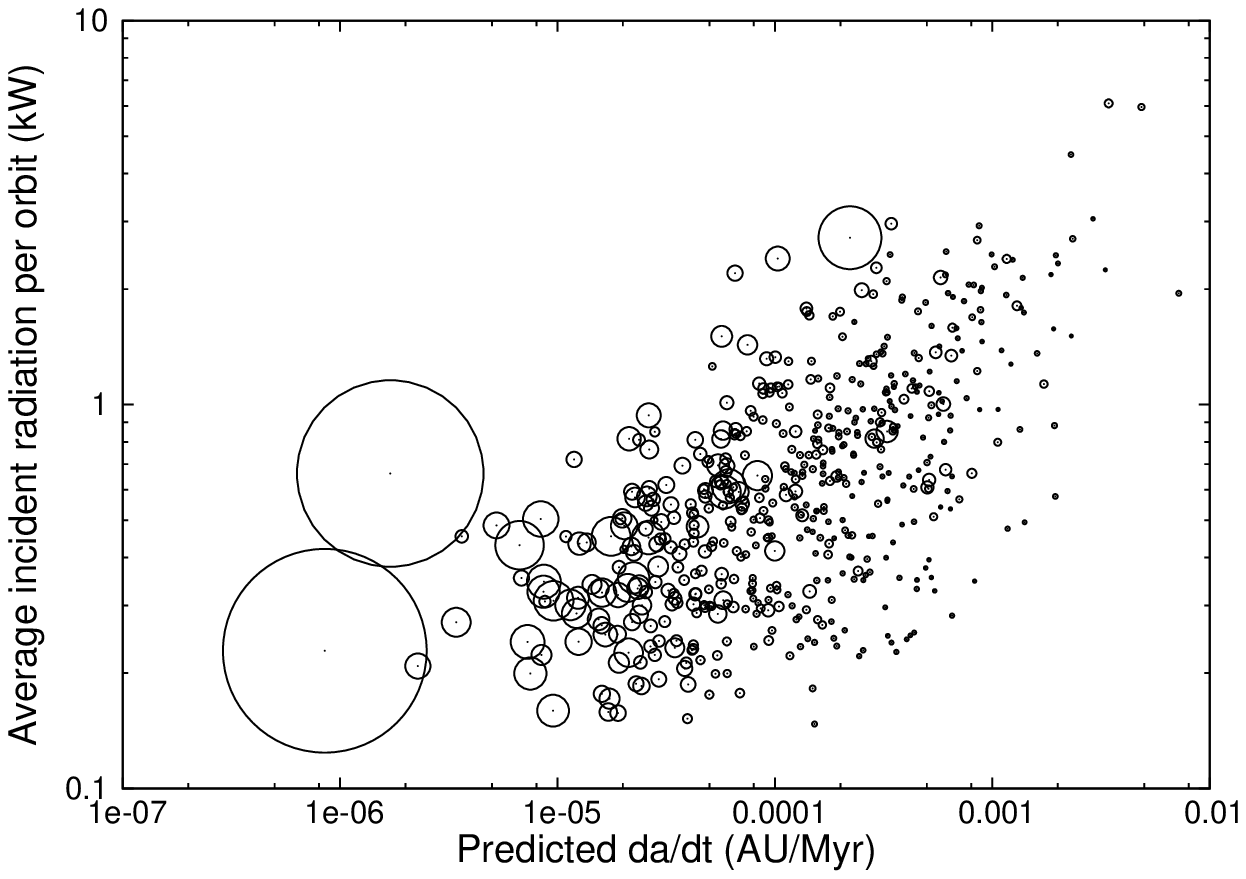}
	\end{center}
	\caption{Relationship between the average incident radiation each NEA receives 
	per orbit and predicted diurnal $da/dt$ drift. Circle sizes are proportional to 
	the diameter of the object. The more sunlight an object receives 
	during its orbit, the more power is available to the Yarkovsky drift. However, 
	this link is tempered by the 
	the diameter-- larger objects experience a smaller drift 
	than smaller objects, given the same average incident radiation.
	\label{fig:radsize}
	}
\end{figure}

The 1/D dependence can be seen in 
Figure \ref{fig:radsize}. As all objects in these plots are 
assumed to have the same bulk density (2000 kg m$^{-3}$), it is only the difference in 
diameters that produces different mass estimates. 

After diameter, the second parameter that strongly influences drift magnitude is 
the average incident radiation per orbit, as seen in Figure \ref{fig:radsize}. The 
more light received by the NEA over its orbit, the more light is available for re-emission and the loss of momentum that powers the drift. 

Many values of $da/dt$ reported in Table \ref{tbl:all} have large error bars due to uncertainties in physical properties. Observations that further constrain the obliquity, density, rotation rate, thermal conductivity and heat capacity would also constrain predicted drift rates. Measurements of thermal properties of these objects would be valuable, as would the measurement of rotation rates and obliquities (either from lightcurves or radar observations). It's expected that 1/6 objects larger than 200 meters are binary systems \citep{MargotBinaryPop,prav06others}, a property which could enable density measurements. 

Historically, Yarkovsky detections require either radar observations over three 
apparitions \citep{ChesleyGolevka} or optical observations that meet a set of 
criteria. \citet{Nugent12} required an object to (1) have an observed arc of at least  $\sim15$ years, 
(2) have observations distributed throughout that arc in time (defined as at least 8 observations per orbit for at least 5 orbits) and (3) have a fraction of 
these observations at favorable geometries and distances (defined by the Yarkovsky sensitivity $s_Y > 2.0$).

None of the objects in Table \ref{tbl:dozen} have enough optical or radar observations to meet the above criteria for detection. Therefore, when possible we 
encourage the community to observe these objects and contribute astrometry to the 
Minor Planet Center. More astrometry is needed for all these objects to enable a future Yarkovsky detection via a fit to optical-only data. 

To facilitate these observations, Table \ref{tbl:obs} provides apparitions and 
associated apparent magnitude ranges for these objects between April 1st, 2012 and 
April 1st, 2022. These apparitions are defined as the times when the object's 
elongation is greater than $90^\circ$, and were generated using the JPL's Horizons 
ephemeris computation service. 

The worldwide community of amateur and professional follow up observers is encouraged to consult this table when planning their observations. Several of the brighter objects may also be automatically picked up by sky surveys such as (in order of decreasing number of observations) PanSTARRS, the Catalina Sky Survey, Spacewatch, and  
the Lincoln Near Earth Asteroid Research Program \citep{Larson07, McMillan07, Stokes00Lincoln}.
 However, some of these objects only have brief windows where their elongation is greater than $90^\circ$ and $V<20.5$ mag (which is roughly the sensitivity limit of most surveys) and may be missed without special attention. The two brightest objects are likely to be automatically observed by surveys, however, additional observations that 
expanded coverage over the orbit in mean anomaly would be useful.

Unfortunately, not all objects are easily observable. 2010 EX11 does not have an elongation greater than $90^\circ$ during that time 
span, though on two apparitions (in 2012 and 2013) it does exceed $60^\circ$.  
Several of the remaining objects are extremely faint, with $V$ (mag) rarely brighter than 23.5. Although these observations may be challenging, they are vital 
for well-defined orbits and future Yarkovsky detections.

\section{Conclusion}
In this paper we use WISE-derived geometric albedos and diameters, as well as values for 
geometric albedos and diameters published in the literature, to produce more accurate diurnal 
Yarkovsky drift predictions for 540 NEAs. Table \ref{tbl:dozen} lists the 12 
objects in our sample with the fastest rates, and Table \ref{tbl:obs} gives their 
apparitions over the next decade. Three of these objects have observed arcs of less 
than a year, and we encourage observers to obtain more astrometry of these objects 
when possible.    
Predicting which NEAs are most likely to be 
subject to strong Yarkovsky drifts relies upon robust determinations of asteroid 
physical properties, underscoring the need to continue to obtain such characterization 
data.



\acknowledgments

This publication makes use of data products from the Wide-field Infrared Survey 
Explorer, which is a joint project of the University of California, Los Angeles, 
and the Jet Propulsion Laboratory/California Institute of Technology, funded by the 
National Aeronautics and Space Administration. This publication also makes use of 
data products from NEOWISE, which is a project of the Jet Propulsion Laboratory/
California Institute of Technology, funded by the Planetary Science Division of the 
National Aeronautics and Space Administration. We gratefully acknowledge the 
extraordinary services specific to NEOWISE contributed by the International 
Astronomical UnionÕs Minor Planet Center, operated by the Harvard-Smithsonian 
Center for Astrophysics, and the Central Bureau for Astronomical Telegrams, 
operated by Harvard University. We also thank the worldwide community of dedicated 
amateur and professional astronomers devoted to minor planet follow-up 
observations. This research has made use of the NASA/IPAC Infrared Science Archive, 
which is operated by the Jet Propulsion Laboratory, California Institute of 
Technology, under contract with the National Aeronautics and Space Administration.

We are very grateful for the work of David Vokrouhlick{\'y}, particularly his 2000 
paper, which is the foundation of this work. We would also wish to thank Jean-Luc
Margot, whose comments markedly improved the manuscript. Finally, we also thank our 
referee for comments that materially improved the quality of this work.

\bibliographystyle{apj}

\begin{thebibliography}{25}
\expandafter\ifx\csname natexlab\endcsname\relax\def\natexlab#1{#1}\fi

\bibitem[{{Bottke} {et~al.}(2006){Bottke}, {Vokrouhlick{\'y}}, {Rubincam}, \&
  {Nesvorn{\'y}}}]{2006AREPS}
{Bottke}, Jr., W.~F., {Vokrouhlick{\'y}}, D., {Rubincam}, D.~P., \&
  {Nesvorn{\'y}}, D. 2006, Annual Review of Earth and Planetary Sciences, 34,
  157

\bibitem[{{Bowell} {et~al.}(1989){Bowell}, {Hapke}, {Domingue}, {Lumme},
  {Peltoniemi}, \& {Harris}}]{bowe89}
{Bowell}, E., {Hapke}, B., {Domingue}, D., {Lumme}, K., {Peltoniemi}, J., \&
  {Harris}, A.~W. 1989, in Asteroids II, 524--556

\bibitem[{{Chamberlin}(2008)}]{SmallBodyDatabase}
{Chamberlin}, A.~B. 2008, Bulletin of the American Astronomical Society, 40

\bibitem[{Chesley {et~al.}({2003})Chesley, Ostro, Vokrouhlicky, Capek,
  Giorgini, Nolan, Margot, Hine, Benner, \& Chamberlin}]{ChesleyGolevka}
Chesley, S., Ostro, S., Vokrouhlicky, D., Capek, D., Giorgini, J., Nolan, M.,
  Margot, J.~L., Hine, A., Benner, L., \& Chamberlin, A. {2003}, {Science},
  {302}, 1739

\bibitem[{{Danby}(1992)}]{Danby}
{Danby}, J.~M.~A. 1992, {Fundamentals of Celestial Mechanics} (Willmann-Bell,
  Inc)

\bibitem[{Giorgini {et~al.}({2002})Giorgini, Ostro, Benner, Chodas, Chesley,
  Hudson, Nolan, Klemola, Standish, Jurgens, Rose, Chamberlain, Yeomans, \&
  Margot}]{Giorgini1950DA}
Giorgini, J., Ostro, S., Benner, L., Chodas, P., Chesley, S., Hudson, R.,
  Nolan, M., Klemola, A., Standish, E., Jurgens, R., Rose, R., Chamberlain, A.,
  Yeomans, D., \& Margot, J. {2002}, {Science}, {296}, 132

\bibitem[{{Larson}(2007)}]{Larson07}
{Larson}, S. 2007, in Near Earth Objects, our Celestial Neighbors: Opportunity
  and Risk, Proceedings if IAU Symposium 236, ed. A.~M. G.~B.~Valsecchi,
  D.~Vokrouhlick{\'y} (Cambridge University Press), 323--328

\bibitem[{{Mainzer} {et~al.}(2011{\natexlab{a}}){Mainzer}, {Bauer}, {Grav},
  {Masiero}, {Cutri}, {Dailey}, {Eisenhardt}, {McMillan}, {Wright}, {Walker},
  {Jedicke}, {Spahr}, {Tholen}, {Alles}, {Beck}, {Brandenburg}, {Conrow},
  {Evans}, {Fowler}, {Jarrett}, {Marsh}, {Masci}, {McCallon}, {Wheelock},
  {Wittman}, {Wyatt}, {DeBaun}, {Elliott}, {Elsbury}, {Gautier}, {Gomillion},
  {Leisawitz}, {Maleszewski}, {Micheli}, \& {Wilkins}}]{Mainzer2011a}
{Mainzer}, A., {Bauer}, J., {Grav}, T., {Masiero}, J., {Cutri}, R.~M.,
  {Dailey}, J., {Eisenhardt}, P., {McMillan}, R.~S., {Wright}, E., {Walker},
  R., {Jedicke}, R., {Spahr}, T., {Tholen}, D., {Alles}, R., {Beck}, R.,
  {Brandenburg}, H., {Conrow}, T., {Evans}, T., {Fowler}, J., {Jarrett}, T.,
  {Marsh}, K., {Masci}, F., {McCallon}, H., {Wheelock}, S., {Wittman}, M.,
  {Wyatt}, P., {DeBaun}, E., {Elliott}, G., {Elsbury}, D., {Gautier}, IV, T.,
  {Gomillion}, S., {Leisawitz}, D., {Maleszewski}, C., {Micheli}, M., \&
  {Wilkins}, A. 2011{\natexlab{a}}, The Astrophysical Journal, 731, 53

\bibitem[{{Mainzer} {et~al.}(2011{\natexlab{b}}){Mainzer}, {Grav}, {Bauer},
  {Masiero}, {McMillan}, {Cutri}, {Walker}, {Wright}, {Eisenhardt}, {Tholen},
  {Spahr}, {Jedicke}, {Denneau}, {DeBaun}, {Elsbury}, {Gautier}, {Gomillion},
  {Hand}, {Mo}, {Watkins}, {Wilkins}, {Bryngelson}, {Del Pino Molina}, {Desai},
  {G{\'o}mez Camus}, {Hidalgo}, {Konstantopoulos}, {Larsen}, {Maleszewski},
  {Malkan}, {Mauduit}, {Mullan}, {Olszewski}, {Pforr}, {Saro}, {Scotti}, \&
  {Wasserman}}]{Mainzer2011d}
{Mainzer}, A., {Grav}, T., {Bauer}, J., {Masiero}, J., {McMillan}, R.~S.,
  {Cutri}, R.~M., {Walker}, R., {Wright}, E., {Eisenhardt}, P., {Tholen},
  D.~J., {Spahr}, T., {Jedicke}, R., {Denneau}, L., {DeBaun}, E., {Elsbury},
  D., {Gautier}, T., {Gomillion}, S., {Hand}, E., {Mo}, W., {Watkins}, J.,
  {Wilkins}, A., {Bryngelson}, G.~L., {Del Pino Molina}, A., {Desai}, S.,
  {G{\'o}mez Camus}, M., {Hidalgo}, S.~L., {Konstantopoulos}, I., {Larsen},
  J.~A., {Maleszewski}, C., {Malkan}, M.~A., {Mauduit}, J.-C., {Mullan}, B.~L.,
  {Olszewski}, E.~W., {Pforr}, J., {Saro}, A., {Scotti}, J.~V., \& {Wasserman},
  L.~H. 2011{\natexlab{b}}, The Astrophysical Journal, 743, 156

\bibitem[{{Mainzer} {et~al.}(2011{\natexlab{c}}){Mainzer}, {Grav}, {Masiero},
  {Bauer}, {Wright}, {Cutri}, {McMillan}, {Cohen}, {Ressler}, \&
  {Eisenhardt}}]{Mainzer2011b}
{Mainzer}, A., {Grav}, T., {Masiero}, J., {Bauer}, J., {Wright}, E., {Cutri},
  R., {McMillan}, R.~S., {Cohen}, M., {Ressler}, M., \& {Eisenhardt}, P.
  2011{\natexlab{c}}, The Astronomical Journal, 736

\bibitem[{{Mainzer} {et~al.}(2011{\natexlab{d}}){Mainzer}, {Grav}, {Masiero},
  {Bauer}, {Wright}, {Cutri}, {Walker}, \& {McMillan}}]{Mainzer2011c}
{Mainzer}, A., {Grav}, T., {Masiero}, J., {Bauer}, J., {Wright}, E., {Cutri},
  R.~M., {Walker}, R., \& {McMillan}, R.~S. 2011{\natexlab{d}}, The
  Astrophysical Journal Letters, 737, L9

\bibitem[{{Margot} {et~al.}(2002){Margot}, {Nolan}, {Benner}, {Ostro},
  {Jurgens}, {Giorgini}, {Slade}, \& {Campbell}}]{MargotBinaryPop}
{Margot}, J.~L., {Nolan}, M.~C., {Benner}, L.~A.~M., {Ostro}, S.~J., {Jurgens},
  R.~F., {Giorgini}, J.~D., {Slade}, M.~A., \& {Campbell}, D.~B. 2002, Science,
  296, 1445

\bibitem[{{McMillan}(2007)}]{McMillan07}
{McMillan}, R.~S. 2007, in Near Earth Objects, our Celestial Neighbors:
  Opportunity and Risk, Proceedings if IAU Symposium 236, ed. A.~M.
  G.~B.~Valsecchi, D.~Vokrouhlick{\'y} (Cambridge University Press), 329--340

\bibitem[{Milani {et~al.}({2009})Milani, Chesley, Sansaturio, Bernardi,
  Valsecchi, \& Arratia}]{MilaniImpact1999RQ36}
Milani, A., Chesley, S.~R., Sansaturio, M.~E., Bernardi, F., Valsecchi, G.~B.,
  \& Arratia, O. {2009}, {Icarus}, {203}, 460

\bibitem[{{Nugent} {et~al.}(2012){Nugent}, {Margot}, {Chesley}, \&
  {Vokrouhlick{\'y}}}]{Nugent12}
{Nugent}, C.~R., {Margot}, J.~L., {Chesley}, S.~R., \& {Vokrouhlick{\'y}}, D.
  2012, accepted

\bibitem[{Opeil {et~al.}(2010)Opeil, Consolmagno, \& Britt}]{OpeilTherCon}
Opeil, C., Consolmagno, G., \& Britt, D. 2010, Icarus, 208, 449

\bibitem[{Pravec \& Harris({2000})}]{PravecFastSlow}
Pravec, P. \& Harris, A. {2000}, {Icarus}, {148}, 12

\bibitem[{{Pravec} {et~al.}(2006)}]{prav06others}
{Pravec}, P. {et~al.} 2006, Icarus, 181, 63

\bibitem[{{Scheeres} {et~al.}(2002){Scheeres}, {Durda}, \&
  {Geissler}}]{ScheeresFateEjecta}
{Scheeres}, D.~J., {Durda}, D.~D., \& {Geissler}, P.~E. 2002, Asteroids III,
  527

\bibitem[{Spitale \& Greenberg({2001})}]{SpitaleGreenberg01}
Spitale, J. \& Greenberg, R. {2001}, {Icarus}, {149}, 222

\bibitem[{{Stokes} {et~al.}(2000){Stokes}, {Evans}, {Viggh}, {Shelly}, \&
  {Pearce}}]{Stokes00Lincoln}
{Stokes}, G.~H., {Evans}, J.~B., {Viggh}, H.~E.~M., {Shelly}, F.~C., \&
  {Pearce}, E.~C. 2000, Icarus, 148, 21

\bibitem[{Vokrouhlick\'{y} {et~al.}({2005})Vokrouhlick\'{y}, Capek, Chesley, \&
  Ostro}]{VokDetect}
Vokrouhlick\'{y}, D., Capek, D., Chesley, S., \& Ostro, S. {2005}, {Icarus},
  {173}, 166

\bibitem[{Vokrouhlick\'{y} {et~al.}({2008})Vokrouhlick\'{y}, Chesley, \&
  Matson}]{Vok1992BF}
Vokrouhlick\'{y}, D., Chesley, S.~R., \& Matson, R.~D. {2008}, {Astronomical
  Journal}, {135}, 2336

\bibitem[{{Vokrouhlick{\'y}} {et~al.}(2000){Vokrouhlick{\'y}}, {Milani}, \&
  {Chesley}}]{VokSmallNEAS}
{Vokrouhlick{\'y}}, D., {Milani}, A., \& {Chesley}, S.~R. 2000, Icarus, 148,
  118

\bibitem[{{Wright} {et~al.}(2010){Wright}, {Eisenhardt}, {Mainzer}, {Ressler},
  {Cutri}, {Jarrett}, {Kirkpatrick}, {Padgett}, {McMillan}, {Skrutskie},
  {Stanford}, {Cohen}, {Walker}, {Mather}, {Leisawitz}, {Gautier}, {McLean},
  {Benford}, {Lonsdale}, {Blain}, {Mendez}, {Irace}, {Duval}, {Liu}, {Royer},
  {Heinrichsen}, {Howard}, {Shannon}, {Kendall}, {Walsh}, {Larsen}, {Cardon},
  {Schick}, {Schwalm}, {Abid}, {Fabinsky}, {Naes}, \& {Tsai}}]{Wright2010}
{Wright}, E.~L., {Eisenhardt}, P.~R.~M., {Mainzer}, A.~K., {Ressler}, M.~E.,
  {Cutri}, R.~M., {Jarrett}, T., {Kirkpatrick}, J.~D., {Padgett}, D.,
  {McMillan}, R.~S., {Skrutskie}, M., {Stanford}, S.~A., {Cohen}, M., {Walker},
  R.~G., {Mather}, J.~C., {Leisawitz}, D., {Gautier}, III, T.~N., {McLean}, I.,
  {Benford}, D., {Lonsdale}, C.~J., {Blain}, A., {Mendez}, B., {Irace}, W.~R.,
  {Duval}, V., {Liu}, F., {Royer}, D., {Heinrichsen}, I., {Howard}, J.,
  {Shannon}, M., {Kendall}, M., {Walsh}, A.~L., {Larsen}, M., {Cardon}, J.~G.,
  {Schick}, S., {Schwalm}, M., {Abid}, M., {Fabinsky}, B., {Naes}, L., \&
  {Tsai}, C.-W. 2010, The Astronomical Journal, 140, 1868

\end{thebibliography}

\clearpage

  \notetoeditor{Table \ref{tbl:all} should appear in online supplementary material.}

\begin{deluxetable}{lrrrrrrrr}
\tabletypesize{\scriptsize}
\tablecaption{The 12 NEAs with largest predicted Yarkovsky drift rates.\label{tbl:dozen}}
\tablewidth{0pt}
\tablehead{
	\colhead{NEA}  &   \colhead{$a$}   &   \colhead{$e$}   &   \colhead{$i$}  &   
\colhead{$D$}   &  \colhead{$p_V$}   &   \colhead{Arc}  &
\colhead{$da/dt$}& \colhead{$\Delta \rho$} \\ 
\colhead{}     &   (AU)   &     &   (deg)   &  (km)   &   &   &    10$^{-4}$ AU/Myr & (km)\\ 
}
\startdata 
                  2010 JG87    & 2.76 &  0.95 & 16.91 &  0.41 $\pm$  0.02 &  0.20 $\pm$  0.04 &   54 days &  72.11 $\pm$ 25.12 &  110.0 \\ 
                  2006 HY51    & 2.60 &  0.97 & 30.58 &  1.22 $\pm$  0.27 &  0.16 $\pm$  0.09 & 2006-2011 &  54.35 $\pm$ 23.85 &   90.8 \\ 
                  2007 EP88    & 0.84 &  0.89 & 20.78 &  0.64 $\pm$  0.04 &  0.17 $\pm$  0.04 & 2007-2010 &  48.57 $\pm$ 16.22 &  443.8 \\ 
         (137924) 2000 BD19    & 0.88 &  0.89 & 25.69 &  0.97 $\pm$  0.04 &  0.25 $\pm$  0.05 & 1997-2010 &  34.30 $\pm$ 12.02 &  292.6 \\   
                  2010 HX107   & 0.80 &  0.30 &  3.37 &  0.06 $\pm$  0.01 &  0.19 $\pm$  0.07 &   60 days &  33.11 $\pm$ 23.05 &  323.6 \\ 
                  2002 LT24    & 0.72 &  0.50 &  0.76 &  0.14 $\pm$  0.02 &  0.14 $\pm$  0.07 & 2002-2010 &  29.06 $\pm$ 16.58 &  333.1 \\ 
         (153201) 2000 WO107   & 0.91 &  0.78 &  7.78 &  0.51 $\pm$  0.08 &  0.13 $\pm$  0.06 & 2000-2010 &  23.45 $\pm$ 10.14 &  188.7 \\   
                  2010 EX11    & 0.96 &  0.11 &  9.75 &  0.04 $\pm$  0.01 &  0.23 $\pm$  0.09 &   37 days &  23.09 $\pm$ 18.64 &  173.0 \\
\enddata
\end{deluxetable}

\begin{deluxetable}{rrrclrr}
\tabletypesize{\scriptsize}
\tablecaption{Observing opportunities for NEAs with highest predicted Yarkovsky drift rates\label{tbl:obs}. 
Apparitions listed are when elongation is greater than $90^\circ$. Apparitions with magnitude ranges always
above 23.5 $V$(mag) are not shown.}
\tablewidth{0pt}
\tablehead{
	\colhead{NEA} & \colhead{$da/dt$}& \colhead{Apparition} & & & \colhead{$V$ (mag) range} & \colhead{Dec range} \\ 
                  & 10$^{-4}$ AU/Myr &                      & & &                           &      degrees        \\
                  }                                                                                           %
\startdata                                                                                                    %
                  2010 JG87    & 64.99 $\pm$ 23.26 &  May 29 2014 & - & Sep 20 2014 & 20.3, 24.4 &  07,  19 \\%
                  2006 HY51    & 34.32 $\pm$ 14.86 &  Feb 08 2014 & - & Jul 06 2014 & 21.5, 23.6 & -12, -03 \\%
                               &                   &  Feb 04 2015 & - & Jun 17 2015 & 20.4, 22.9 &  06,  22 \\%
                               &                   &  Feb 05 2018 & - & Jul 08 2018 & 22.3, 24.0 & -10, -02 \\%
                               &                   &  Mar 11 2019 & - & Jun 05 2019 & 17.9, 21.8 &  24,  53 \\%
                               &                   &  Jan 25 2020 & - & Jul 01 2020 & 23.4, 24.6 &  00,  08 \\%
                               &                   &  Feb 03 2022 & - & Apr 01 2022 & 23.3, 24.3 & -09, -06 \\%
                  2007 EP88    & 31.04 $\pm$ 12.04 &  Mar 14 2013 & - & May 14 2013 & 17.2, 21.2 & -58, -03 \\
\enddata
\end{deluxetable}

 \begin{deluxetable}{lrrrrrrrr}
\tabletypesize{\scriptsize}
\tablecaption{NEAs with highest predicted Yarkovsky drift rates\label{tbl:all}}
\tablewidth{0pt}
\tablehead{
	\colhead{NEA}  &   \colhead{$a$}   &   \colhead{$e$}   &   \colhead{$i$}  &   
\colhead{$D$}   &  \colhead{$p_V$}   &   \colhead{Arc}  &
\colhead{$da/dt$}& \colhead{$\Delta \rho$} \\ 
\colhead{}     &   (AU)   &     &   (deg)   &  (km)   &   &   &    10$^{-4}$ AU/Myr & (km)\\ 
}
\startdata 
                  2010 JG87    & 2.76 &  0.95 & 16.91 &  0.41 $\pm$  0.02 &  0.20 $\pm$  0.04 &   54 days &  72.11 $\pm$ 25.12 &  110.0 \\ 
                  2006 HY51    & 2.60 &  0.97 & 30.58 &  1.22 $\pm$  0.27 &  0.16 $\pm$  0.09 & 2006-2011 &  54.35 $\pm$ 23.85 &   90.8 \\ 
                  2007 EP88    & 0.84 &  0.89 & 20.78 &  0.64 $\pm$  0.04 &  0.17 $\pm$  0.04 & 2007-2010 &  48.57 $\pm$ 16.22 &  443.8 \\ 
         (137924) 2000 BD19    & 0.88 &  0.89 & 25.69 &  0.97 $\pm$  0.04 &  0.25 $\pm$  0.05 & 1997-2010 &  34.30 $\pm$ 12.02 &  292.6 \\   
                  2010 HX107   & 0.80 &  0.30 &  3.37 &  0.06 $\pm$  0.01 &  0.19 $\pm$  0.07 &   60 days &  33.11 $\pm$ 23.05 &  323.6 \\ 
                  2002 LT24    & 0.72 &  0.50 &  0.76 &  0.14 $\pm$  0.02 &  0.14 $\pm$  0.07 & 2002-2010 &  29.06 $\pm$ 16.58 &  333.1 \\ 
         (153201) 2000 WO107   & 0.91 &  0.78 &  7.78 &  0.51 $\pm$  0.08 &  0.13 $\pm$  0.06 & 2000-2010 &  23.45 $\pm$ 10.14 &  188.7 \\   
                  2010 EX11    & 0.96 &  0.11 &  9.75 &  0.04 $\pm$  0.01 &  0.23 $\pm$  0.09 &   37 days &  23.09 $\pm$ 18.64 &  173.0 \\
\enddata
\end{deluxetable}

\end{document}